\documentclass[12pt]{article}

\usepackage{amsmath,amssymb,dsfont}
\usepackage{color}
\usepackage{epsfig}

\usepackage[text={15.9cm,21.7cm}]{geometry}

\begin{document}

\title{
\vspace{-3.0cm} 
{\normalsize
\hfill\mbox{}TUM-HEP 809/11\\}
\vspace{2.5cm} 
 \mbox{Understanding neutrino properties from decoupling}
\mbox{right-handed neutrinos and extra Higgs doublets}}

\author{Alejandro Ibarra and Cristoforo Simonetto \\
\\ \footnotesize
Physik-Department T30d, Technische Universit\"at M\"unchen,
\\ \footnotesize
James-Franck-Stra\ss{}e, 85748 Garching, Germany}

\date{}

\maketitle

\begin{abstract}

Low energy effects induced by heavy extra degrees of freedom are suppressed
by powers of the large mass scale, thus preserving, if sufficiently
heavy, the successes of the Standard Model in describing low energy 
phenomena. However, as is well known, heavy right-handed neutrinos may play
an important role in low energy phenomenology as an explanation
of the smallness of neutrino masses. We consider in this paper an
extension of the Standard Model by heavy right-handed neutrinos and
heavy Higgs doublets and we show, 
using a renormalization group approach,
that this model can simultaneously provide an explanation for
the small neutrino masses and for the mild hierarchy observed between
the atmospheric and the solar mass splittings, even when the extra
degrees of freedom are very heavy. 
We analyze the necessary conditions
to reproduce the oscillation parameters and we discuss possible
experimental signatures of this model.

\end{abstract}

\section{Introduction}

The last fifteen years have witnessed a tremendous experimental 
progress in neutrino physics, leading to a good
determination of two mass splittings, two mixing angles and
a fairly stringent bound on the third angle~\cite{GonzalezGarcia:2007ib}.
The most conspicuous difference between quark
and neutrino properties is without any doubt the smallness of 
neutrino masses. Furthermore, the better and better measurements of neutrino
parameters have also revealed the existence of large mixing
angles in the leptonic sector and the existence of a 
relatively mild mass hierarchy between the two heaviest neutrino 
masses~\cite{Maltoni:2008ka}.

Extending the particle content of the Standard Model (SM) with three heavy 
right-handed  neutrinos, thus implementing the (type I) see-saw 
mechanism~\cite{seesaw}, 
solves very elegantly the problem of generating small neutrino
masses and opens new opportunities to understand the puzzles of the
existence of large mixing angles and a small mass hierarchy. Unfortunately,
whereas
the seesaw mechanism does not make any generic prediction about the
leptonic mixing matrix, it tends to predict a neutrino mass hierarchy
which is much larger than the one inferred from 
experiments~\cite{Casas:2006hf}. 
Namely, under the
plausible assumption that the neutrino Yukawa couplings have hierarchical
eigenvalues, as observed in the quark and the charged lepton sectors,
the mild mass hierarchy observed between the solar and the atmospheric 
mass splitting can only be accommodated in very special cases. One
possibility arises when the hierarchies between the masses of the 
heavy right-handed neutrinos is much larger than the hierarchy 
between the neutrino Yukawa eigenvalues, in which 
case the right-handed mixing angles have to be tiny. 
Alternatively, when the hierarchy in the masses of the heavy right-handed
neutrinos is comparable to the hierarchy in the Yukawa couplings,
it is possible to accommodate the observed mild neutrino mass hierarchy
only for certain, possibly fine-tuned, choices of the 
right-handed mixing angles.

Further extensions of the type I see-saw model have been considered
in the literature. A minimal possibility 
consists on introducing one extra Higgs doublet,
which leads to radiatively induced neutrino masses. Most works impose
an additional $Z_2$ symmetry, under which the Standard Model particles
are even, whereas the right-handed neutrinos and the extra Higgs
doublet are odd~\cite{Ma:2006km,Z2}.
With this assignment, the tree level neutrino mass
vanishes and the leading contribution is hence the radiatively generated
one. In this scenario, the right-handed neutrinos and the extra Higgses
could be directly produced at colliders while being the neutrino
masses in the measured range, thanks to the loop suppression and 
an appropriate choice of the neutrino Yukawa and Higgs self-couplings.
Furthermore, the $Z_2$ symmetry ensures the absence of tree level 
flavour changing neutral currents. On the other hand, it has also
been discussed the general two Higgs doublet model (2HDM), without 
imposing ad-hoc discrete symmetries~\cite{Grimus:1989pu}.
In this case, both mechanisms of neutrino 
mass generation are present, naturally leading to a mild neutrino
mass hierarchy~\cite{Grimus:1999wm}. 

In this paper we carefully analyze the mechanism of neutrino mass generation
in the two Higgs doublet model extended with right-handed neutrinos.
Rather than being motivated by finding signatures of new physics in experiments
at the energy frontier or at the intensity frontier, we are motivated by
constructing a simple and natural framework capable to explain the observed
neutrino parameters while preserving the successes of the
Standard Model. This approach is inspired in the high scale
see-saw mechanism, which despite its well known lack of
testability, still stands as the most compelling explanation for the
small neutrino masses.

We will argue that by making all the new particles
heavy it is possible to simultaneously explain the smallness of
neutrino masses and the mildness of the neutrino mass hierarchy,
without jeopardizing any of the successes of the Standard Model.
Remarkably, in this model only
one right-handed neutrino suffices to generate two neutrino mass
scales: the atmospheric neutrino mass scale
will be generated at tree level, whereas the solar mass scale will be 
generated by the radiative corrections to the effective neutrino mass matrix.
Both neutrino masses are suppressed
by the large right-handed neutrino mass scale, thus explaining
the tininess of neutrino masses. On the other hand, the ratio 
of the two mass scales is suppressed by the loop
factor and enhanced by a large logarithm of the ratio of the heavy 
right-handed neutrino mass to the Higgs mass, resulting in a neutrino mass
ratio which can be roughly of the correct size.
Finally, the decoupling of the extra Higgs degrees of
freedom ensures the absence of large contributions to the
flavour and CP violating processes, both in the leptonic sector
and in the quark sector, as well as to the electroweak precision
measurements.

In Section \ref{sec:twoHDM} we review the basic features of the
two Higgs doublet model, the various problems which arise in 
this very minimal extension of the Standard Model, and how
they can be circumvented altogether by decoupling the extra
scalar degrees of freedom. In Section \ref{sec:numasses}
we show, using a renormalization group approach, that 
even in the decoupling limit
the extra Higgs particles can play an important role 
in low energy neutrino physics, as an explanation for
the mild hierarchy observed between the atmospheric and the
solar neutrino mass scales. The viability of this model
requires at least one right-handed neutrino and two Higgs doublets;
in Section \ref{sec:2hdm2rhn} we comment on the differences of this
framework with
another minimal framework of neutrino masses, namely 
the two right-handed neutrino model with just one Higgs doublet.
In Section \ref{sec:corrections} we calculate the corrections to the
leptonic mixing matrix induced by quantum effects, and we 
argue that a non-zero $\theta_{13}$ is generically expected, as well
as a a deviation from the maximal atmospheric angle which is
correlated to the angle $\theta_{13}$. Lastly, in Section 
\ref{sec:conclusions} we present our conclusions.

\section{Benefits of the decoupling limit of the 2HDM}\label{sec:twoHDM}

We consider an extension of the SM consisting on adding to the 
particle content one additional Higgs doublet, with identical quantum numbers 
as the SM Higgs doublet. The most general Higgs potential 
reads~\cite{Lee:1973iz,Branco:2011iw}:
\begin{align}\label{eq:potential}
V&= m_{11}^2\Phi_1^\dagger\Phi_1+m_{22}^2\Phi_2^\dagger\Phi_2
-[m_{12}^2\Phi_1^\dagger\Phi_2+{\rm h.c.}]\nonumber\\
& +\frac{1}{2}\lambda_1(\Phi_1^\dagger\Phi_1)^2
+\frac{1}{2}\lambda_2(\Phi_2^\dagger\Phi_2)^2
+\lambda_3(\Phi_1^\dagger\Phi_1)(\Phi_2^\dagger\Phi_2)
+\lambda_4(\Phi_1^\dagger\Phi_2)(\Phi_2^\dagger\Phi_1)
\nonumber\\
& +\left[\frac{1}{2}\lambda_5(\Phi_1^\dagger\Phi_2)^2
+\lambda_6(\Phi_1^\dagger\Phi_1)(\Phi_1^\dagger\Phi_2)
+\lambda_7(\Phi_2^\dagger\Phi_2)(\Phi_1^\dagger\Phi_2)+
{\rm h.c.}\right]\,.
\end{align}

Despite being such a simple extension of the Standard Model,
the introduction of a second Higgs doublet in general jeopardizes
many of the successes of the Standard Model. More concretely,
the Higgs potential has now a richer structure including 
electrically charged directions, which may lead to the 
spontaneous breaking of the electromagnetic $U(1)$ symmetry
if there are minima along those directions. Besides, 
the extra Higgs doublet contributes to the oblique parameters
$S$, $T$ and $U$~\cite{Peskin:1990zt}, 
possibly leading to values in conflict with 
electroweak precision data. Lastly, the
new, in general flavour violating, couplings of the fermions
to the extra Higgs doublet could lead to too large flavour
changing neutral currents or lepton flavour violation. 

It is interesting that all these problems are simultaneously
solved in the decoupling limit of the two Higgs doublet model,
which is defined as the limit where one of the Higgses is
kept light, with a mass comparable to the Z-boson mass, 
while the rest acquire masses much larger than $M_Z$.

To show the absence of charge breaking minima in the decoupling limit
we will work without loss of generality in the Higgs basis where $m_{12}=0$.
Then, in complete analogy to the Standard Model,
we postulate the absence of unbounded from below directions
in the potential.
Furthermore, and also in analogy to the Standard Model,
we require that one mass squared, say $m_{11}^2$, is negative,
to allow the spontaneous breaking of the electroweak symmetry.
Lastly, a necessary condition for the decoupling of the second Higgs is
$m^2_{22}\gg |m^2_{11}|/\lambda_1>0$~\cite{Gunion:2002zf}.

To study the minima of the potential, we will exploit the 
$SU(2)_L$ invariance to express the Higgs fields as
\begin{align}
\Phi_1=\frac{1}{\sqrt{2}}\begin{pmatrix} 0 \\ \varphi_1 \end{pmatrix}~~~~~~
\Phi_2=\frac{1}{\sqrt{2}}\begin{pmatrix} \sigma \\ \varphi_2 \end{pmatrix}\;,
\end{align}
where $\sigma$ and $\varphi_2$ are complex fields, while
$\varphi_1$ is a real field.

We consider now the direction in field space
$\varphi_2 = a \varphi_1$, $\sigma = b \varphi_1$, along which 
the potential reads:
\begin{equation}\label{eq:Vab}
 V_{a,b}(\varphi_1) =\frac{1}{2} m_{11}^2 |\varphi_1|^2 +
                    \frac{1}{2}m_{22}^2 (|a|^2+|b|^2) |\varphi_1|^2 
                    + {\rm const.}\times |\varphi_1|^4\;.
\end{equation}
Clearly, for any minimum of $V$ with vacuum expectation values
$\left< \varphi_1 \right>$, 
$\left< \varphi_2 \right>$, $\left< \sigma\right>$
there is a choice of the parameters $a$ and $b$ such that 
$V_{a,b}(\varphi_1)$ has a minimum at $\left< \varphi_1 \right>$, concretely
for $a= \left< \varphi_2 \right>/ \left< \varphi_1 \right>$ and 
$b= \left< \sigma \right>/ \left< \varphi_1 \right>$.
This is only possible, though, 
if the quadratic part of eq.~\eqref{eq:Vab} is negative,
thus for any minimum of $V$ it must hold that:
\begin{equation}
 |\left< \varphi_2 \right>|^2 + |\left< \sigma\right>|^2 <
    \frac{|\langle\varphi_1\rangle|^2}{m_{22}^2} |m_{11}^2|\;.
\end{equation}

Utilizing this inequality it is straightforward to determine the minimum of $V$.
The differentiation with respect to $\varphi_1$ yields
\begin{equation}
|\langle\varphi_1\rangle|^2=\frac{2|m_{11}^2|}{\lambda_1}
  \left[1+{\cal O}\left(\sqrt{\frac{|m_{11}^2|}{\lambda_1 m_{22}^2}}\right)
\right]\;.
\end{equation}
Varying now $V$ with respect to $\sigma$ and $\varphi_2$, 
it can be checked that 
\begin{align}
\begin{split}
& \langle\sigma\rangle = 0 \;,\\
&\langle \varphi_2\rangle\simeq-\langle\varphi_1\rangle \frac{\lambda_6^* \langle\varphi_1\rangle ^2}{2 m_{22}^2}\;,
\end{split}
\end{align}
which shows that the electric charge is conserved. Furthermore,
as the decoupling limit is approached, the Standard Model vacuum is
recovered.

Apart from recovering the nice features of the Standard Model vacuum,
when taking the decoupling limit also the successful predictions of the
Standard Model in electroweak observables and flavour physics are 
recovered. Namely, the existence of an additional Higgs doublet introduces
contributions to the oblique parameters $S$, $T$ and $U$
which can be in tension with the electroweak precision measurements. 
However, it can be shown that in the decoupling limit $S$, $T$ and $U$
scale with $|m_{11}^2|/(\lambda_1 m_{22}^2)$~\cite{Haber:2010bw}, 
thus bringing the oblique parameters within their experimentally 
allowed values for sufficiently heavy extra scalar states.

Besides, the general 2HDM  induces in general too large rates for the
flavour changing neutral currents and the lepton flavour violating processes.
Whereas this problem can be alleviated by assuming concrete flavour
structures of the Yukawa couplings \cite{alignment}, a simpler way to suppress
altogether all new contributions to the flavour violating processes
consists on assuming that the new scalar particles are all very heavy.

The general flavour dependent part of the Lagrangian reads
\begin{align}\label{eq:YukawaSM}
-{\cal L}^{\rm Yuk}
& = 
(Y_e^a)_{ij} \bar l_{Li} e_{Rj} \Phi_a + (Y_u^a)_{ij} \bar q_{Li} u_{Rj} \tilde \Phi_a+
(Y_d^a)_{ij} \bar q_{Li} d_{Rj} \Phi_a+{\rm h.c.}
\end{align}
where $i,j=1,2,3$  are flavour indices, $a=1,2$ is a Higgs index
and $\tilde{\Phi}_a=i\tau_2\Phi^*_a$.
It will be convenient in what follows to work in the Higgs basis
where one of the Higgs fields, say $\Phi_2$, does not acquire a
vacuum expectation value. In this basis, then, the Yukawa matrices
$Y_{e,u,d}^1$ are proportional to the fermion mass matrices.

Consider for illustration the contribution of the second Higgs doublet
to the process $\mu\rightarrow e\gamma$, whose rate is strongly
constrained by experiments $\mathrm{BR} (\mu \rightarrow e\, \gamma) < 1.2\times 10^{-11}$~\cite{Brooks:1999pu}. For a wide range of parameters, this process is dominated by the 
two loop Barr-Zee diagrams~\cite{Paradisi:2005tk}. The leading contribution comes from the top quark unless  $Y_{u33}^2$ is small \cite{Hisano:2010es}.
In the decoupling limit the branching ratio reads:
\begin{align}
  \label{eq:BRmuegamma}
 \mathrm{BR} (\mu \rightarrow e\, \gamma) &\simeq
  \frac{8 \alpha^3}{3 \pi^3}
   \frac{|Y_{e12}^2|^2}{|Y_{e22}^1|^2}
  \left| f \!\! \left(\frac{m_t^2}{m_h^2} \right) \cos \alpha-
  \frac{Y_{u33}^2}{Y_{u33}^1} \frac{m_t^2}{m_H^2} \log^2 \frac{m_t^2}{m_H^2} \right|^2\;.
\end{align}
Here, $m_t$ denotes the top quark mass, $f(z)$ is 
defined in \cite{Hisano:2010es} and evaluates $f(2)\approx 1$, and
$\alpha$ is the Higgs mixing angle, which reads 
 $\cos \alpha \simeq|\lambda_6|v^2/m_H^2$ in the decoupling limit.
Thus, the stringent experimental bound on $\mathrm{BR} (\mu \rightarrow e\, \gamma)$ can be evaded, for the concrete flavour structure $Y_{e12}^2 = \sqrt{Y_{e11}^1 Y_{e22}^1}$ and $Y_{u33}^2 = Y_{u33}^1$, if the scale of the heavy Higgs $m_H \gtrsim 2~\mathrm{TeV}$. Clearly, the rate of $\mu\rightarrow e\gamma$ can always be suppressed, regardless of the flavour structure of the Yukawa couplings, by increasing sufficiently the mass of the extra scalar degrees of freedom.

In the quark sector, strong constraints come from the measurement of the
mass difference in meson anti-meson systems, such as
$B^0_s$ -- $\bar B^0_s$, which arise at tree level in the general 2HDM.
In the decoupling limit, the $B^0_s$ -- $\bar B^0_s$ mass difference 
approximately reads:
\begin{equation}
 \Delta m_{B_s} \simeq \left|\Delta m_{B_s}^{\rm SM}+
\frac{4}{3} m_{B_s} f_{B_s}^2 P_2^{LR} \frac{Y^{2*}_{d23} Y^{2}_{d32}}{m_H^2}\right|\;.
\end{equation}
Where $f_B$ is the $B$-meson decay constant,
 $f_{B_s}=238.8\pm 9.5 {\rm ~MeV}$~\cite{Laiho:2009eu}, $m_{B_s}$ is the 
$B_s$ meson mass, $m_{B_s}=5.37 $ GeV~\cite{Nakamura:2010zzi},
and the coefficient $P_2^{LR}$ includes the renormalization group 
evolution from the scale $M_Z$ to $\sim m_{B_S}$ and the hadronization 
of the quarks to mesons~\cite{Buras:2001ra} and reads $P_2^{LR}\simeq 3.0$.
Assuming $|Y^{2*}_{d23}|\approx| Y^{2}_{d32}| \approx \sqrt{Y^{1}_{d22} Y^{1}_{d33}}$,
we estimate that the  2HDM contribution to the 
$B^0_s$ -- $\bar B^0_s$ mass difference lies within the 
theoretical error of the SM calculation, 
$\Delta m_{B_s}^{\rm SM}=(135\pm 20)\times 10^{-13}$ GeV~\cite{Lunghi:2007ak},
for $m_H \gtrsim 3~\mathrm{TeV}$. As before, the 2HDM contribution
to the meson-antimeson mixing can always be suppressed for a sufficiently
large $m_H$, regardless of the flavour structure of the quark 
Yukawa couplings.

\section{Neutrino masses in a 2HDM extended with right-handed neutrinos}\label{sec:numasses}

We will consider in this paper an extension of the Standard Model 
consisting in adding one extra Higgs doublet and at least one right-handed
neutrino, singlet under the Standard Model gauge group. We
will not impose any discrete symmetry on the model. Then, compatible
with this matter content, the most general Lagrangian reads:
\begin{align}
{\cal L}={\cal L}^{\rm kin}+{\cal L}^{\rm Yuk}+{\cal L}^{\nu}-V\;,
\end{align}
where ${\cal L}^{\rm kin}$ contains the kinetic terms, 
${\cal L}^{\rm Yuk}$ is the Yukawa Lagrangian for the Standard
Model fermions, given in eq.~(\ref{eq:YukawaSM}), 
$V$ is the Higgs potential, given in eq.~(\ref{eq:potential})
and ${\cal L}^\nu$ is the part of the Lagrangian involving
right-handed neutrinos, given by:
\begin{equation}
-{\cal L}^{\nu}=
(Y_\nu^a)_{ij} \bar l_{Li} \nu_{Rj} \tilde \Phi_a 
-\frac{1}{2} {M_{\rm M}}_{ij} \bar\nu^C_{Ri}\nu_{Rj}+{\rm h.c.}
\end{equation}
This term contains a Yukawa coupling, which leads to Dirac neutrino
masses, and a Majorana mass for the right-handed neutrinos, with
a size which is a priori unrelated to the electroweak symmetry
breaking scale.

We will assume that the 
mass scale of the right-handed neutrinos is much larger
than the electroweak symmetry breaking scale and the mass of all the
extra Higgs mass eigenstates $H^0$, $A^0$, $H^\pm$, which we denote
collectively by $m_H$.
Hence,  the right-handed neutrinos are decoupled,
leading to the following effective operators:
\begin{align}
-{\cal L}^{\nu,~\rm eff} &=
+\frac{1}{2}\kappa_{ij}^{ab} 
(\bar l_{Li} \tilde \Phi_a) 
(\tilde \Phi_b^T  l^C_{Lj}) +{\rm h.c.}
\end{align}
where, at the scale of the lightest right-handed neutrino,
\begin{align}
\kappa^{ab}(M_1)=(Y_\nu^a M_{\rm M}^{-1} Y_\nu^{b\;T})(M_1)\;.
\label{eq:kappaab}
\end{align}

Since we have chosen to work in the basis where 
$\langle \Phi^0_1\rangle=v/\sqrt{2}$, $\langle \Phi^0_2 \rangle=0$, 
the neutrino mass matrix at the scale of the lightest right-handed
neutrino depends just on the coupling $\kappa^{11}$:
\begin{align}
{\cal M}_\nu(M_1)=\frac{v^2}{2}\kappa^{11}(M_1)\;,
\end{align}
which is diagonalized in the standard way:
\begin{align}
{\cal M}_\nu=U^* {\rm diag}(m_1,m_2,m_3)U^\dagger\;.
\end{align}
These are not, however, the neutrino parameters 
measured by experiments, where the energies involved are much smaller
than the right-handed Majorana mass scale. 
In order to compare the predictions of the model with low energy
experiments we will make use of the Renormalization Group Equations 
(RGE) given in the Appendix, to run the effective couplings
$\kappa^{ab}$ from the scale $M_1$ to the scale $m_H$.
Below the scale $m_H$ the neutrino mass matrix
runs with the RGEs of the Standard Model extended with massive 
Majorana neutrinos. Since in this framework the neutrino masses
are hierarchical, the running will not introduce any new
qualitative feature but will only modify the values of the mass
eigenvalues and the entries of the leptonic mixing matrix 
by a small factor, proportional to the tau Yukawa coupling squared and 
to the small logarithm $\log(m_H/M_Z)$~\cite{Casas:1999tg}.

To emphasize the main features of the quantum corrections to
the neutrino mass matrix, we will 
concentrate in what follows on a see-saw model with just one 
right-handed neutrino with mass $M_{\rm maj}$. 
Then, with this assumption, the neutrino Yukawa couplings $Y_{\nu}^a$ are
3-vectors. In this scenario
the neutrino mass matrix at the scale of the right-handed neutrino
mass $M_{\rm maj}$ is given by:
\begin{align}
[\kappa^{11}]^{\rm tree}=\frac{Y_{\nu}^{1}  Y_{\nu}^{1\;T}}{M_{\rm maj}}\;,
\end{align}
which has rank 1 and thus only one non-vanishing eigenvalue.

On the other hand, quantum effects introduce corrections to
the neutrino mass matrix yielding at low energies
 $\kappa^{11}(m_H)=[\kappa^{11}]^{\rm tree}+\delta\kappa^{11}$,
where the correction reads, in the leading-log approximation,
\begin{align}
\delta\kappa^{11}=-\frac{1}{16\pi^2}\beta_{\kappa^{11}}
(M_{\rm maj})\log\frac{M_{\rm maj}}{m_H}\;.
\end{align}

Using the explicit form of the $\beta$ function in the Appendix,
it follows that this correction can be schematically
written as:
\begin{align}
\delta \kappa^{11}\simeq
B_{1a}\kappa^{a1}+\kappa^{1a}B_{1a}^T+ b \kappa^{22}\;.
\label{correction}
\end{align}
Here $B_{1a}$ denote 
$3\times 3$ matrices whereas $b$ is a number.
The first two terms generalize the well known correction to the neutrino
mass matrix in the Standard Model including the dimension-5 Weinberg
operator. However, the term proportional to $\kappa^{22}$ does
not have any correspondence in the Standard Model and, as we will see,
introduces new qualitative features. The coefficient $b$ explicitly
reads:
\begin{align}
b=-\frac{1}{16\pi^2} 2 \lambda_5\log\frac{M_{\rm maj}}{m_H}\;,
\end{align}
which depends linearly on the coefficient of the potential
term  $\lambda_5 (\Phi_1^\dagger \Phi_2)(\Phi_1^\dagger \Phi_2)$,
while only logarithmically on the ratio between the
scale of the right-handed neutrino
and the overall scale of the extra scalars $H^0$, $H^\pm$, $A^0$.

The neutrino mass matrix ${\cal M}_\nu = [{\cal M}_\nu]^{\rm tree}+
\delta {\cal M}_\nu$ can be diagonalized using perturbation theory,
giving as a result the eigenvalues $m_i=m_i^{(0)}+\delta m_i$ with
$m_i^{(0)}$ the eigenvalue at tree level and $\delta m_i$ 
the first order correction.

At lowest order in perturbation theory, taking
into account only the tree level mass term, there is only one
non-vanishing neutrino mass eigenvalue:
\begin{align}\label{eq:m3}
m_3^{(0)}=\frac{v^2}{2 M_{\rm maj}}  |Y^1_\nu|^2\;.
\end{align}
On the other hand, the third column of the leptonic mixing matrix reads;
\begin{align}
U^{(0)}_{i3}=\frac{Y^{1*}_{\nu i}}{|Y^1_\nu|}\;,
\label{U0i3}
\end{align}
while the first two columns are undefined, due to the degeneracy of
the corresponding neutrino mass eigenvalues. In this expression,
$|Y^a_\nu|=(\sum_i |Y^a_{\nu i}|^2)^{1/2}$.

The correction to the neutrino mass eigenvalues due to the perturbation
$\delta\kappa^{11}$ is given by:
\begin{align}
\delta m_i=\frac{v^2}{2}{\rm Re}[(U^{(0)T}\,\delta\kappa^{11} 
\, U^{(0)})_{ii}]\;,
\end{align}
which slightly modifies the value of the 
heaviest neutrino mass eigenvalue:
\begin{align}
 \delta m_3 = \frac{v^2}{2 M_{\rm maj}}  {\rm Re} \left[ 
	 2(Y_\nu^{1\dagger} B_{1a} Y_\nu^a)
         +b \frac{(Y_\nu^{1\dagger} Y_\nu^2)^2}{{|Y_\nu^1|}^2}\right] \;.
\label{eq:m3corr}
\end{align}

More importantly, this correction is
also non-vanishing for $\delta m_2$, thus generating radiatively a 
second neutrino mass eigenvalue. This is in contrast to the widely
studied case of the Standard Model extended with a single right-handed
neutrino, where there is only one non-vanishing neutrino mass eigenvalue,
even after taking into account the renormalization group running.
\footnote{There are, however, tiny finite corrections 
arising from two-loop diagrams involving W bosons~\cite{Petcov:1984nz}.}

To show this, we write explicitly the radiative correction to the 
next-to-lightest neutrino mass eigenvalue:
\begin{align}
\delta m_2=\frac{v^2}{2}{\rm Re}[U^{(0)}_{p2}\,\delta\kappa^{11}_{pq} 
\, U^{(0)}_{q2}]\;,
\end{align}
which crucially depends on the second column of the zero-th order
leptonic mixing matrix.
Since the matrix $U^{(0)}$ is unitary, the vector $U^{(0)}_{q2}$ 
should satisfy $\sum_q U^{(0)}_{q2} U^{(0)*}_{q3}=0$,
$\sum_q U^{(0)}_{q2} U^{(0)*}_{q2}=1$. A vector that satisfies those properties
can be easily constructed from the vectors $Y^1_\nu$ and $Y^2_\nu$
using the Gram-Schmidt process. Starting with $U^{(0)}_{i3}$ given by 
eq.~(\ref{U0i3}) one finds:
\begin{align}
U^{(0)}_{i2}=\frac{1}{N_2}\left[
Y^{2*}_{\nu i}-\frac{ Y^{2\dagger}_\nu Y^1_\nu }{|Y^1_\nu|} 
\frac{Y^{1*}_{\nu i}}{|Y^1_\nu|}\right]
{e^{-\frac{i}{2} \arg (-\lambda_5)} }\;,
\label{U0i2}
\end{align}
where
\begin{align}
N_2=\left[ Y_\nu^{2\dagger} Y_\nu^2 -
\frac{{| Y^{2\dagger}_\nu Y^1_\nu|}^2}{{|Y^1_\nu|}^2}\right]^{1/2}\;.
\end{align}
Substituting into the expression for $\delta m_2$ we find
\begin{align}\label{eq:m2}
m_2= \frac{1}{16\pi^2}\frac{|\lambda_5| v^2}{M_{\rm maj}}
\left[{| Y_\nu^2|}^2- \frac{{| Y^{2\dagger}_\nu Y^1_\nu|}^2}{{|Y^1_\nu|}^2}\right]\log\frac{M_{\rm maj}}{m_H}\;.
\end{align}
(Note that the phase in eq.~(\ref{U0i2}) has been chosen to
yield $m_2$ real and positive.)
It is apparent from this expression that 
in order to generate a non-vanishing neutrino
mass eigenvalue it is necessary the misalignment between the Yukawa couplings
$Y^1_\nu$ and $Y^2_\nu$, or in more physical terms, it
is necessary the existence
of new sources of flavour violation in the neutrino sector. 
These new sources necessarily generate, through quantum corrections,
off-diagonal elements in the charged lepton Yukawa coupling  $Y^2_e$, which
in turn induce a contribution to the lepton flavour violating processes.
Nevertheless, as explained in Section \ref{sec:twoHDM} 
this contribution is suppressed by the large mass of the extra 
Higgs particles, and can be consistent with experiments 
if the extra particles are sufficiently heavy.

Furthermore, it is interesting to note that, under some well motivated
assumptions, the hierarchy between the tree level mass $m_3$ and the 
radiatively generated neutrino mass $m_2$ can be fairly mild. 
For instance, taking the typical values $|\lambda_5|\sim 1$, 
$M_{\rm maj}\sim 10^{11}\,{\rm GeV}$ and $m_H\sim 1\,{\rm TeV}$
and assuming non-aligned neutrino Yukawa couplings with 
$|Y^2_\nu|\sim |Y^1_\nu|$ one obtains for the ratio between the two heaviest
neutrino mass eigenvalues:
\begin{align}
\frac{m_2}{m_3}\simeq \frac{|\lambda_5|}{8\pi^2}
\frac{| Y^{2}_\nu|^2}{{|Y^1_\nu|}^2}
\log\frac{M_{\rm maj}}{m_H}
\sim 0.2\;,
\label{mildhierarchy}
\end{align}
which yields a mild mass hierarchy, in qualitative
agreement with the experimental data. Note that, whereas the overall
scale of the light neutrino masses depends linearly on the inverse
of the heavy right-handed neutrino mass, the ratio between the two
heaviest neutrino mass eigenvalues depends only logarithmically with 
the masses of the new particles. As a consequence, the result in
eq.~(\ref{mildhierarchy}) is fairly insensitive to the exact 
values of the masses of the heavy particles.

In the previous analysis we have assumed for simplicity
that only one right-handed neutrino participates in the
neutrino mass generation. In the more realistic case
where there are several right-handed neutrinos, the tree level
contributions to all neutrino mass eigenvalues will be non-vanishing.
Nevertheless, as discussed in the Introduction, if the neutrino Yukawa coupling
$Y_\nu^1$ has hierarchical eigenvalues, as suggested by the observed
hierarchies in the quark and charged lepton masses, then the 
neutrino mass hierarchy generated (at tree level) by the see-saw mechanism is
in general several orders of magnitude larger than the one inferred
from experiments. Therefore, the radiatively generated contribution 
to the next-to-lightest neutrino masses by the presence of the
second Higgs doublet will dominate over the tree level contribution,
and the conclusions presented above will still hold.

In an extended scenario with $N_H$ Higgs doublets and one right-handed
neutrino, the radiatively induced
next-to-lightest neutrino mass receives $N_H \times (N_H-1)/2$ contributions,
each of them proportional to the coefficient of the term 
$\lambda_5^{ab}(\Phi^\dagger_a \Phi_1)(\Phi^\dagger_b \Phi_1)$, $a,b=2...N_H$,
in the Higgs potential. Therefore, in this case the radiatively
generated neutrino mass is enhanced. 

It is amusing to speculate that adding more Higgs
doublets to the particle content of the
model may also be relevant to understand the observed pattern of
neutrino mixing angles. In a model with $N_H$ Higgs doublets, only
the Higgs that acquires a vacuum expectation value, $\Phi_1$, will
contribute to the tree level mass. Assuming that this is the
largest mass,  $m_3=\sqrt{\Delta m^2_{\rm atm}}$, it follows from 
eq.~(\ref{U0i3}) that $U_{i3}\propto Y_{\nu i}^{1*}$. Therefore, if there
is any pattern in the neutrino Yukawa coupling $ Y_{\nu i}^{1}$, stemming e.g.
from an underlying flavour symmetry, then this pattern will be
inherited by $U_{i3}$, thus providing an explanation to the 
apparent structure of the last column of the leptonic mixing angle:
$|U_{13}|\simeq 0$, $|U_{23}|\simeq |U_{33}|$. In contrast,
there are $N_H-1$ Higgses which contribute via quantum effects
to the generation of the solar neutrino mass scale,  $m_2=\sqrt{\Delta m^2_{\rm sol}}$ and of the
second column of the leptonic mixing matrix, $U_{i2}$. As a consequence, even
if there is a structure in each of the neutrino Yukawa couplings, 
$Y_{\nu i}^{a}$, $a=2, ...N_H$, the generated $U_{i2}$ will be structureless,
since it receives contributions from all these Yukawa couplings. This
is in rough agreement with observations, which reveal that the three
entries in $U_{i2}$ are all ${\cal O}(0.1)$, without displaying
any remarkable structure (or, alternatively, the solar angle is
neither maximal nor zero). Therefore, in the Standard
Model extended with right-handed neutrinos and several Higgs doublets,
the last column of the leptonic mixing matrix is expected to display
a ``hierarchical'' structure, whereas the second column, an ``anarchical''
structure~\cite{Hall:1999sn}, in qualitative agreement with the data.

\section{Comparison to the two right-handed neutrino\\ model}\label{sec:2hdm2rhn}

The scenario discussed in this paper leads to a dimension-5 operator
which is identical to the one generated by the Standard Model
(with a single Higgs doublet) extended by two heavy right-handed neutrinos.
There are however some conceptual differences in the way these two
scenarios reproduce the observed neutrino data, which we discuss here.

Let us first demonstrate the equivalence of the two Higgs doublet
model extended with one right-handed neutrino (2HD-1RHN model) and
the Standard model extended with two right-handed neutrinos (1HD-2RHN model).
At low energies, the coefficient of the dimension five operator
generated in the 2HD-1RHN model reads, following 
eqs.~(\ref{eq:kappaab}), (\ref{correction}),
\begin{align}
 \kappa^{11}(m_H) &\simeq 
\frac{1}{M_{\rm maj}}\left[
 Y_\nu^1 {Y_\nu^1}^T+ B_{1a} Y_\nu^a {Y_\nu^1}^T+Y_\nu^1 {Y_\nu^a}^TB_{1a}^T+ b Y_\nu^2 {Y_\nu^2}^T\right] \;,
\end{align}
which can be recast as
\begin{align}\label{eq:kappaBB}
\kappa^{11}(m_H) \simeq 
\frac{1}{M_{\rm maj}}\left[
 (Y_\nu^1 \! +\! B_{1a} Y_\nu^a)(Y_\nu^1 \! +\! B_{1a} Y_\nu^a)^T- B_{1a} Y_\nu^a {Y_\nu^{a'}}^TB_{1a'}^T +  b Y_\nu^2 {Y_\nu^2}^T\right] \;.
\end{align}
Neglecting the term of ${\cal O}(B^2)$ and defining 
$\tilde Y_\nu^1=Y_\nu^1 + B_{1a} Y_\nu^a$, one obtains the following low
energy neutrino mass matrix
\begin{align}
{\cal M}_\nu &\simeq \left(\frac{\tilde Y_\nu^1 {\tilde Y_\nu}^{1\,T}}{M_{\rm maj}}
+\frac{b Y_\nu^2 {Y_\nu^2}^T}{M_{\rm maj}}\right) \frac{v^2}{2}\;.
\label{eq:numass2HDM}
\end{align}
This result is formally identical to the 
effective operator which arises in the low energy
limit of the 1HD-2RHN model. More specifically, in this model
the high energy Lagrangian reads, in the basis where the $2\times 2$ 
right-handed neutrino mass matrix is diagonal,
\begin{equation}
-{\cal L}^{\nu, \rm 2RHN}=
(Y_\nu)_{ij} \bar l_{Li} \nu_{Rj} \tilde \Phi
-\frac{1}{2} M_1 \bar\nu^C_{R1}\nu_{R1}
-\frac{1}{2} M_2 \bar\nu^C_{R2}\nu_{R2}+{\rm h.c.}\;,
\end{equation}
which leads, when $M_1,~M_2\gg v$, to the effective neutrino mass matrix
\begin{align}
{\cal M}_\nu^{\rm 2RHN} &\simeq \left(\frac{Y_1 Y_1^T}{M_1}
+\frac{ Y_2 Y_2^T}{M_2}\right) \frac{v^2}{2}\;,
\label{eq:numass2RHN}
\end{align}
being $Y_{1,2}$ column vectors defined as $Y_{1}\equiv (Y_\nu)_{i1}$,
$Y_{2}\equiv (Y_\nu)_{i2}$.
By comparing eqs.~(\ref{eq:numass2HDM}) and (\ref{eq:numass2RHN}) 
it follows that, from the point of view of the neutrino mass generation,
the 2HD-1RHN model is equivalent to the 1HD-2RHN model
with the following correspondence among parameters
\begin{align}
\{Y_1,Y_2,M_1,M_2\}\leftrightarrow 
\{\tilde Y^1_\nu, Y^2_\nu, M_{\rm maj}, M_{\rm maj}/b\}\;.
\end{align}

This correspondence allows to write explicit expressions 
for the most general Yukawa 
couplings $Y_\nu^1$,  $Y_\nu^2$ which lead to the neutrino
masses $m_2, m_3$ and the leptonic mixing matrix $U$.
Using the results of \cite{Casas:2001sr}, one easily finds:
\begin{align}
{\tilde Y}^1_\nu&=\frac{\sqrt 2}{v}\sqrt{M_{\rm maj}}
(\sqrt{m_2} \cos\hat\theta U_{i2}^* \pm
\sqrt{m_3}\sin\hat\theta U_{i3}^* )\;,\\
Y^2_\nu&=\frac{\sqrt 2}{v}\sqrt{\frac{M_{\rm maj}}{b}}
(-\sqrt{m_2}\sin\hat\theta U_{i2}^*\pm 
\sqrt{m_3}\cos\hat\theta U_{i3}^*)\;,
\end{align}
where $\hat \theta$ is a complex angle which parametrizes 
the family of Yukawa couplings compatible with the low energy
neutrino data (note that this parametrization may fail 
if the term of ${\cal O}(B^2)$ in eq.~(\ref{eq:kappaBB}) can not be neglected).
Finally, the Yukawa coupling with the Higgs 
$\Phi_1$ is $ Y_\nu^1= (1- B_{11}) \tilde Y_\nu^1- B_{12} Y_\nu^2$.

Furthermore, this correspondence allows to better appreciate the advantages
of the 2HD-1RHN model over the 1HD-2RHN model in the generation of a 
mild mass hierarchy. In the 1HD-2RHN model, the neutrino mass hierarchy
is essentially given by
\begin{align}
\frac{m_3}{m_2}\sim \frac{|Y_2|^2}{|Y_1|^2}\frac{M_1}{M_2}\;,
\end{align}
therefore the only possibilities to generate a mild neutrino
mass hierarchy are i) $|Y_2|^2\sim |Y_1|^2$, $M_2\sim M_1$, or
ii) $|Y_2|\gg |Y_1|$ with $M_2/ M_1 \sim |Y_1|^2/|Y_2|^2$.
In view of the observed large hierarchies in the quark 
and charged lepton Yukawa eigenvalues, in a model with two
right-handed neutrinos one expects $|Y_2|\gg |Y_1|$,
which hence requires a huge hierarchy between the two right-handed
neutrino masses in order to render a mild light neutrino mass hierarchy.
More concretely, if the neutrino Yukawa eigenvalues have a similar 
hierarchy as the up-type quark masses, $|Y_2|/|Y_1|\sim m_t/m_c\sim
150$, then it is required $M_2/M_1\sim 20000$. On the other
hand, if the hierarchy is similar to the down-type quark masses,
$|Y_2|/|Y_1|\sim m_b/m_s\sim 40$, then it is required 
$M_2/M_1\sim 1600$. Whereas such large hierarchies in the right-handed
neutrino masses cannot be precluded, it is difficult to 
conceive that in the decoupling limit the large hierarchies 
in the Yukawa couplings 
cancel almost exactly a huge hierarchy in the right-handed neutrino
masses to generate at low energies a light
neutrino mass hierarchy of $\sim 6$, as inferred from experiments.

This drawback is very naturally circumvented by the 2HD-1RHN model which 
as we argued above is equivalent, concerning the neutrino mass generation, 
to the 1HD-2RHN model. 
In the equivalent 1HD-2RHN model, the Yukawa couplings $Y_1$
and $Y_2$ are naturally of the same size, since in
the original 2HD-1RHN they correspond to 
Yukawa couplings to the {\it same} generation of right-handed
neutrinos. Furthermore, in the corresponding two right-handed
neutrino model, the masses $M_1$ and $M_2$ naturally present 
a mild hierarchy, given by the factor $1/b\sim{\cal O}(1-10)$.
As a result, the 2HD-1RHN is equivalent to a 1HD-2RHN model
which naturally fulfills the conditions i) to reproduce
the observed mild neutrino mass hierarchy.

Another important difference between the 1HD-2RHN model and
the 2HD-1RHN model concerns the possibility of observing 
other phenomena at low energies apart from neutrino masses.
It is well known that, in its simplest version, the 1HD-2RHN model
does not have any other observable low energy prediction apart
from the tininess of neutrino masses.
In this model the scale of lepton flavour and lepton number violation 
both coincide with the scale of the right-handed Majorana neutrino
masses, which are postulated to be much larger than the electroweak
symmetry breaking scale. As a consequence, the rates of all 
flavour and lepton number violating processes are inversely
proportional to the heavy right-handed neutrino mass
resulting in tiny rates.\footnote{A notable exception is the 1HD-2RHN scenario where
the right-handed neutrinos form a pseudo-Dirac pair with masses 
${\cal O}(100-1000)$ GeV. In this case, the Yukawa couplings can
be sizable while correctly reproducing the tininess of the neutrino
masses. As a consequence, the rates for $\mu\rightarrow e \gamma$ and 
neutrinoless double beta decay can be largely enhanced, 
possibly allowing their observation in the next round of
experiments~\cite{Ibarra:2011xn}.} 
In contrast, in the 2HD-1RHN model, apart from the lepton flavour
associated to the right-handed neutrino couplings there exist
another source of lepton flavour violation associated to the 
charged lepton couplings to the second Higgs doublet, inducing
rates for the rare lepton decays suppressed by the heavy Higgs
masses. If the additional scalar degrees of freedom have masses
not far from the electroweak symmetry breaking scale, the 
induced rates of the lepton flavour violating processes could be large
enough to be observed in experiments. 
A more detailed discussion about
the prospects to observe the process $\mu\rightarrow e\gamma$ in 
experiments will be presented in Section \ref{sec:lfv}.

\section{Corrections to the mixing angles and discussion of $\boldsymbol{\sin \theta_{13}}$}
\label{sec:corrections}

Below the right-handed neutrino mass scale, 
the neutrino mixing angles receive radiative corrections 
with two different origins. First, the change in the entries
of the neutrino mass matrix due to the RGE running, eq.~(\ref{correction}),
generates a correction to the leptonic mixing matrix given by:
\begin{align}
\delta U_\kappa=U^{(0)} T\;,
\end{align}
where
\begin{align}
T_{ii}&\equiv-\frac{i [U^{(0)\,T}\delta {\cal M}_\nu\, U^{(0)}]_{ii}}{2m^{(0)}_i} \;, \\
T_{ij}&\equiv
\frac{m^{(0)}_i [U^{(0)\,T}\delta {\cal M}_\nu\, U^{(0)}]_{ij}
+m^{(0)}_j [U^{(0)\,T}\delta {\cal M}_\nu\, U^{(0)}]_{ij}^*}
{m^{(0)\,2}_j-m^{(0)\,2}_i}
~~{\rm if~}i\neq j \;.
\end{align}

However, this is not the physical leptonic mixing matrix measured
by experiments, since the RGE running also modifies the structure of
the charged lepton Yukawa couplings. More specifically, if the charged
lepton Yukawa coupling $Y_e^1$ is diagonal at $M_{\rm maj}$, the radiative
corrections induced by $Y_e^2$ will generate at low energies off-diagonal
entries in $Y_e^1$. It is then necessary to redefine the charged lepton
fields in order to render a diagonal charged lepton Yukawa coupling,
namely $l_{L}\rightarrow V_e^L l_{L}$, $e_R\rightarrow V_e^R e_R$, where  
 $V_e^L$, $V_e^R$ follow from the singular value decomposition, 
$Y_e^1=V_e^L {\rm diag} (y_{e1}^1, y_{e2}^1, y_{e3}^1) V_e^{R\dagger}$.
This redefinition introduces an additional correction to the leptonic
mixing matrix given by
\begin{align}
\delta U_{Y_e}=(V_e^L-\mathds{1})^T U^{(0)}\;.
\end{align}

The matrix  $V_e^L$ can be explicitly calculated
from the $\beta$-functions of the charged lepton Yukawa couplings.
Using
\begin{equation}\label{eq:YeYe}
 \left. Y_e^1 Y_e^{1\dagger} \right|_{m_H} =
 Y_e^1 Y_e^{1\dagger} -\frac{1}{16\pi^2}(\beta_{Y_e^1} Y_e^{1\dagger} +
Y_e^1 \beta_{Y_e^1}^\dagger)
\log\frac{M_{\rm maj}}{m_H}\;,
\end{equation}
we obtain
\begin{equation}\label{eq:VeL}
(V_e^L)_{ij}=-\frac{1}{16\pi^2}\frac{(\beta_{Y_e^1} Y_e^{1\dagger} +
Y_e^1 \beta_{Y_e^1}^\dagger)_{ij}}{ (y_{ej}^1)^2-(y_{ei}^1)^2}
\log\frac{M_{\rm maj}}{m_H}~~~~i\neq j\;.
\end{equation}

Therefore, summing up the two contributions, the leptonic mixing matrix
at low energies reads, in the physical basis,
\begin{align}
U^{(1)}=V_e^{L\,T} U^{(0)}+U^{(0)}T\;.
\end{align}

We are particularly interested in the correction to the last column
of the leptonic mixing matrix, which in general yields a non-vanishing
contribution to $\sin\theta_{13}$ and a deviation from the maximal 
atmospheric mixing which may be observed in experiments.
Concretely, the correction to the third column of the leptonic mixing matrix
due to the running of $\kappa$ reads
\begin{align} \label{eq:deltaUkappa}
 (\delta U_\kappa)_{i3} = (U^{(0)}T)_{i3} &= \frac{Y_{\nu i}^{1*}}{|Y_\nu^1|} \left[
     -\frac{{\rm Re} (Y_\nu^{1\dagger} B_{1a} Y_\nu^a)}{|Y_\nu^1|^2}
        + \frac{i}{2} \frac{{\rm Im} (b^* (Y_\nu^{2\dagger} Y_\nu^1)^2)}{|Y_\nu^1|^4}\right]\nonumber\\
&\quad + \frac{(B_{1a}^* Y_\nu^{a*})_i}{|Y_\nu^1|}
    +\left( Y_{\nu i}^{2*} - Y_{\nu i}^{1*} \frac{(Y_\nu^{2\dagger} Y_\nu^1)}{|Y_\nu^1|^2} \right) b^* \frac{Y_\nu^{2\dagger} Y_\nu^1}{|Y_\nu^1|^3}\;,
\end{align} 
while the contribution from the rediagonalization of the 
charged lepton Yukawa coupling reads:

\begin{align}
\label{eq:deltaUY}
( \delta U_{Y_e})_{i3} =
- \frac{1}{16 \pi^2} \sum_{j\neq i}
\frac{(\beta_{Y_e^1} Y_e^{1\dagger} +
Y_e^1 \beta_{Y_e^1}^\dagger)_{ji}}{ (y_{ei}^1)^2-(y_{ej}^1)^2}
\frac{ Y_{\nu\,j}^{1*}}{|Y_\nu^{1}|}
\log\frac{M_{\rm maj}}{m_H}\;.
\end{align}

A quantity of particular interest is the angle $\theta_{13}$,
which is constrained by present experiments to be small.
It is interesting that radiative corrections can generate 
in this model a fairly large value of $\theta_{13}$, possibly
at the reach of the planned experiments, even if its tree-level
value vanishes. Summing up the contributions
from eqs.~(\ref{eq:deltaUkappa}) and (\ref{eq:deltaUY}), and neglecting
terms cubic in the charged lepton Yukawa couplings,
 we obtain that the radiatively induced value of $U_{13}$ is
\begin{align}\label{eq:deltaU13}
\delta U_{13} &= -\frac{1}{16\pi^2} \frac{Y_{\nu1}^{2*}}{|Y_\nu^1|}	\left[ 
        	3{\rm Tr}(Y^{1\dagger}_{u}Y^{2}_{u}+ Y^{1}_{d}Y^{2\dagger}_{d})
		+2\lambda_6^* + 2\lambda_5^* \frac{Y_\nu^{2\dagger} Y_\nu^1}{|Y_\nu^1|^2}
	\right]    \log{\frac{M_{\rm maj}}{m_H}} \nonumber\\
&\quad +\frac{1}{16\pi^2} \frac{(Y_\nu^{1\dagger} (Y^1_e)^{-1} Y^{2\dagger}_e)_1}{|Y_\nu^1|}
	\left[
		3 {\rm Tr}(Y^{2\dagger}_{u}Y^{1}_{u}+Y^{2}_{d}Y^{1\dagger}_{d})
	\right] \log\frac{M_{\rm maj}}{m_H}\;,
\end{align} 
which is, as the ratio $m_2/m_3$, suppressed by
the loop factor but enhanced by the large logarithm of the 
ratio of the Majorana mass over the heavy Higgs mass.
As a result, the radiatively generated $\theta_{13}$ can
be as large as $\sim 0.2$ if any of the entries in the 
bracket is $\sim {\cal O}(1)$. 

So far we have considered only the corrections to the neutrino
mixing angles from the running below the right-handed
neutrino mass scale. However, in a large class of models the
cut-off of the theory lies at higher energies and additional
contributions to the leptonic mixing may arise from radiative corrections
between the cut-off scale $\Lambda$ and the right-handed neutrino mass
scale $M_{\rm maj}$. The corrections for this case can be derived using
eqs.~(\ref{eq:deltaUkappa}), (\ref{eq:deltaUY}) and the substitution rules given in Appendix~\ref{sec:aboveMmaj}, the result being:
 \begin{align}\label{eq:deltaU13Lambda}
 \delta U_{13} &= -\frac{Y_{\nu1}^{2*}}{|Y_\nu^1|}
  \bigg\{\!\!\begin{aligned}[t]
        &\left[{\rm Tr}(3Y^{1\dagger}_{u}Y^{2}_{u} + 3Y^{1}_{d}Y^{2\dagger}_{d}+Y^{1\dagger}_{\nu}Y^2_{\nu}) +2 Y^{1\dagger}_\nu (Y^1_e)^{-1}Y^{2\dagger}_e Y^{1}_\nu \right]
\frac{\log\frac{\Lambda}{M_{\rm maj}}}{16\pi^2} \\ 
&+\left[3{\rm Tr}(Y^{1\dagger}_{u}Y^{2}_{u}+ Y^{1}_{d}Y^{2\dagger}_{d})+
2\lambda_6^* + 2\lambda_5^* \frac{Y_\nu^{2\dagger} Y_\nu^1}{|Y_\nu^1|^2} \right]
      \frac{\log{\frac{M_{\rm maj}}{m_H}}}{16\pi^2}
    \bigg\} \end{aligned} \nonumber\\
 &\quad+ \frac{(Y_\nu^{1\dagger} (Y^1_e)^{-1} Y^{2\dagger}_e)_1}{|Y_\nu^1|}
  \Big\{
    {\rm Tr}(Y^{2\dagger}_{\nu}Y^{1}_{\nu}) \frac{\log\frac{\Lambda}{M_{\rm maj}}}{16\pi^2}
    +3 {\rm Tr}(Y^{2\dagger}_{u}Y^{1}_{u}+Y^{2}_{d}Y^{1\dagger}_{d}) \frac{\log\frac{\Lambda}{m_H}}{16\pi^2}
    \Big\} \;.
 \end{align} 

Quantum effects also induce corrections to the
atmospheric mixing angle, leading to deviations to the maximal
mixing even if $\theta_{23}=\pi/4$ at tree level. 
It is interesting that if the 
neutrino Yukawa couplings are the dominant source of flavour violation 
in the leptonic sector, then a correlation arises between the deviations
of $U_{23}/U_{33}$ and $U_{13}$ from their corresponding values at the cut-off scale. 

In this limit, the radiative corrections to the last column of the 
leptonic mixing matrix are dominated by
the first line of eq.~(\ref{eq:deltaU13}), which can be schematically written as:
\begin{equation}
  U_{i3} = (1+\epsilon_3) U^{(0)}_{i3} + \epsilon_2 U^{(0)}_{i2}\;.
\end{equation}
It can be checked that to first order the ratio $U_{23}/U_{33}$ does not depend on 
$\epsilon_3$. Then, using the equation for  $U_{13}$ to eliminate $\epsilon_{2}$
it follows that:
\begin{equation}\label{eq:correlation-master}
  \frac{U_{23}}{U_{33}}-\frac{U^{(0)}_{23}}{U^{(0)}_{33}}\simeq
  \frac{U^{(0)}_{22}U^{(0)}_{33} - U^{(0)}_{32} U^{(0)}_{23} }{U^{(0)2}_{33}}\,
  \frac{U_{13}-U^{(0)}_{13}}{U^{(0)}_{12}}\;.
\end{equation}

Concretely, in the case when at the cut-off scale
the atmospheric mixing angle is exactly maximal and $\theta_{13}$ vanishes,
at low energies the elements of the leptonic mixing matrix 
approximately satisfy
\begin{equation}
  \frac{U_{23}}{U_{33}}-1
  \simeq 2 \sqrt{2} U_{13}\;,
\end{equation}
which can be recast as
\begin{align}\label{eq:correlation}
 \tan \theta_{23} &\simeq |1+ 2 \sqrt{2} \sin \theta_{13} e^{-i\delta}|
  & \text{or} &&
  \theta_{23} -\frac{\pi}{4} &\simeq
\sqrt{2} \sin\theta_{13} \cos{\delta}\;.
\end{align}

\begin{figure}
\hspace{-3mm}
 \epsfig{figure=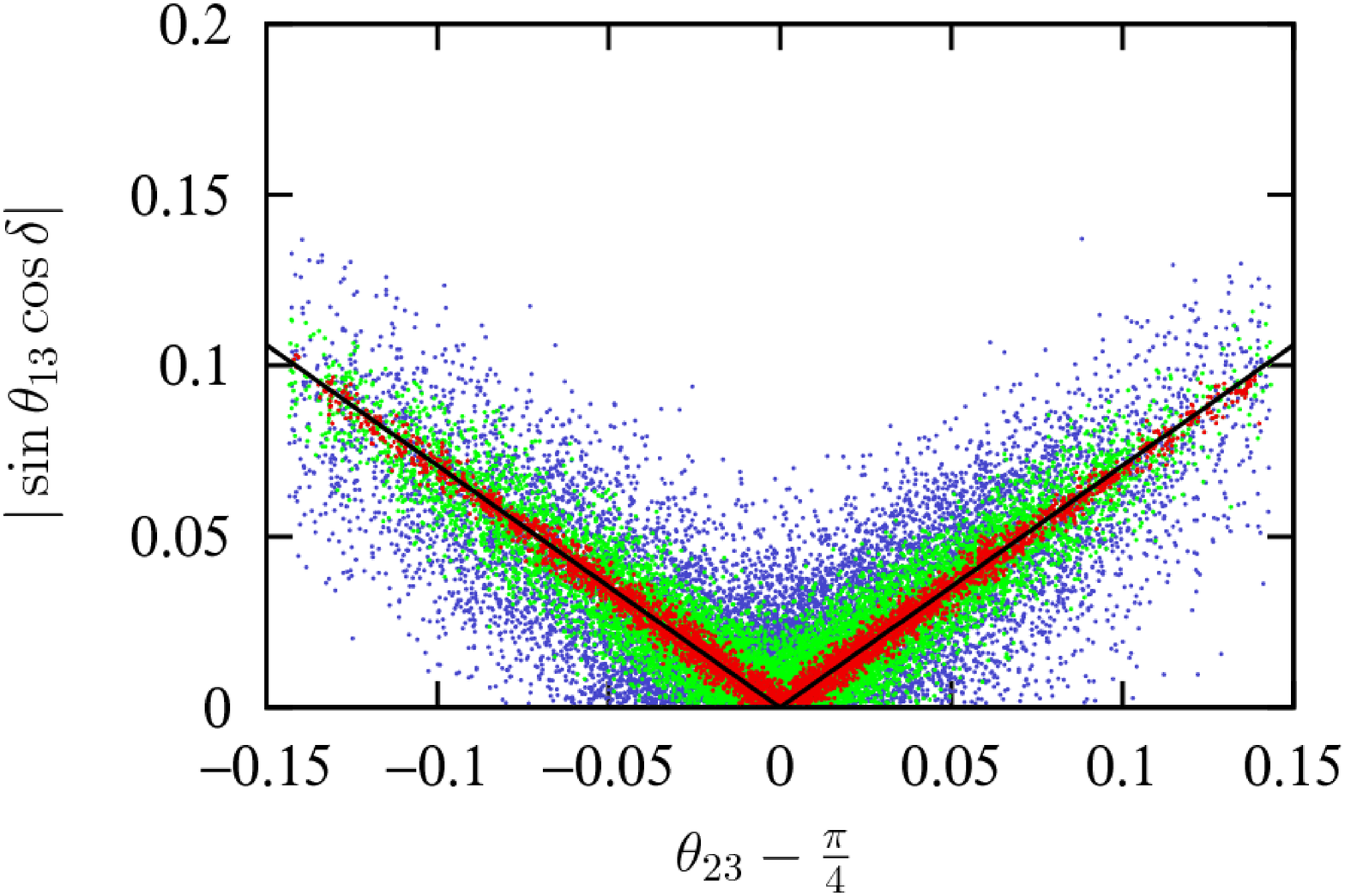,width=80mm}
 \epsfig{figure=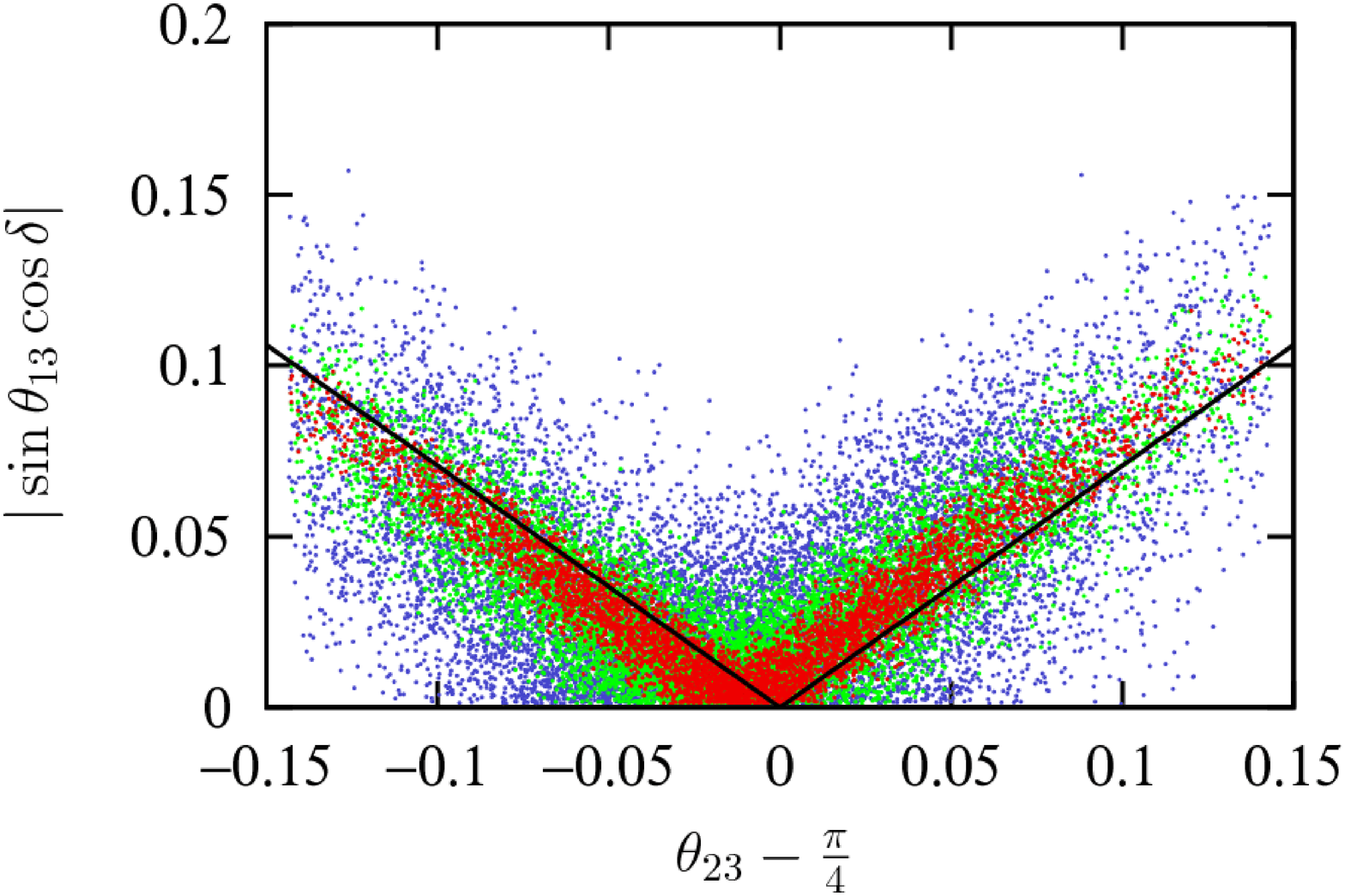,width=80mm}
 \caption{Scatter plots showing $|\sin \theta_{13} \cos \delta |$ against 
  $\theta_{23}-\frac{\pi}{4}$ at low energies for random choices
  of high energy parameters consistent with the measured neutrino oscillation
  parameters. We have assumed 
  tri-bi-maximal mixing at the cut-off scale, being the deviation
  from $\theta_{23}=\pi/4$ and $\theta_{13}=0$ at low energies 
  only due to the radiative corrections
  as described in the main text. The cut-off scale is 
  $\Lambda= M_{\rm maj} = 10^{14} {\rm ~GeV}$ in the left panel and 
  $\Lambda=10^{18} {\rm ~GeV}$ in the right panel.}
  \label{fig:correlation}
\end{figure}

If there are additional sources of lepton flavour violation, 
then the low energy
values of $\theta_{23}$ and $\theta_{13}$ are expected to deviate from 
this relation. 
This is illustrated in the scatter plots shown in fig.~\ref{fig:correlation},
which have been obtained by the numerical one loop integration of the RGEs
of the 2HDM extended by one right-handed neutrino. We assume in the plot
$m_H = 3 {\rm ~TeV}$, $M_{\rm maj} = 10^{14} {\rm ~GeV}$ and 
tri-bi-maximal mixing at a cut-off scale, which we
take $\Lambda=M_{\rm maj}$ ($\Lambda=10^{18} {\rm ~GeV}$) in the left (right)
panel. Quantum effects generate a non-vanishing value for $\theta_{13}$
and  $\theta_{23}-\frac{\pi}{4}$, mostly due to the RGE effects of
$\lambda_5$ and $\lambda_6$, as follows from 
eqs.~(\ref{eq:deltaU13}) and (\ref{eq:deltaU13Lambda}).
In the plot we fix $|\lambda_5|=0.5$ and we take random values
with $|\lambda_6|<0.45$, in order to preserve the perturbativity
of the quartic couplings in the renormalization group running.
To investigate the impact of the charged lepton mixing in
the correlation we have adopted the ansatz $Y_e^2=V Y_e^1$, where $V$ is a
general unitary matrix with random angles and phases. Furthermore, since the
effect of the charged lepton Yukawa couplings on the corrections to the 
leptonic mixing matrix is proportional to ${\rm Tr} (Y_u^1 Y_u^{2\dagger})$, 
we have taken in the scatter plot $|(Y^2_u)_{33}/ (Y^1_u)_{33}| \leq 0.05$ 
(red points), $| (Y^2_u)_{33}/ (Y^1_u)_{33}| \leq 0.15$ (green points) and
 $| (Y^2_u)_{33}/ (Y^1_u)_{33}| \leq 0.3$ (blue points); 
the effects of the down quark Yukawa couplings have been neglected in 
this analysis, although their role is completely analogous.
All the points in the plot reproduce the neutrino oscillation parameters within
their experimental errors.

It is apparent from the plots that when the charged lepton Yukawa couplings
have a negligible effect on the running  (corresponding to
$|(Y^2_u)_{33}/ (Y^1_u)_{33}| \ll 1$), there is a fairly strong correlation
between the radiatively generated $\theta_{13}$ and $\theta_{23}-\frac{\pi}{4}$. 
When the cut-off is $\Lambda=M_{\rm maj}$, the numerical results are
in good agreement with eq.~\eqref{eq:correlation}, shown as a black
solid line in the plot. In contrast, when $\Lambda=10^{18} {\rm ~GeV}$ there 
is a larger spread of the points, due to the additional RGE effects between
$\Lambda$ and $M_{\rm maj}$. Besides, in this case the numerical results
do not agree with eq.~\eqref{eq:correlation}, since the RGE running
between $\Lambda$ and $M_{\rm maj}$ generates a non-vanishing (and negative) 
shift of $U_{23}/U_{33}$ at the scale $M_{\rm maj}$, even if $U_{13}$ still 
vanishes. This produces, following eq.~\eqref{eq:correlation-master},
the shift of the points to the left of the black solid line.

To summarize, from our analytical and numerical analysis, it follows that in the 2HDM extended
with one right-handed neutrino it is generally expected a deviation of the atmospheric
angle from the maximal value which is comparable to the reactor angle
\begin{equation}
\left|\theta_{23}-\frac{\pi}{4}\right|\approx \theta_{13}\;.
\end{equation}
unless the CP violating phase $\delta$ is very close to $\pi/2$.

\section{Lepton flavour violation}\label{sec:lfv}

In the general 2HDM extended with RH neutrinos, one generically
expects a misalignment in the charged lepton Yukawa couplings which
will lead to new phenomena at low energies, apart from neutrino
masses, in contrast to the standard see-saw 
scenario with just one Higgs doublet. This misalignment will generically
arise already at tree level. However, even if the charged lepton
Yukawa couplings are aligned at the cut-off scale $\Lambda$,
radiative corrections from the neutrino Yukawa couplings from the
RGE running above the Majorana mass scale will introduce off-diagonal
entries in both charged lepton Yukawa matrices. Note that the radiative
generation of neutrino masses {\it requires} a misalignment in the neutrino
Yukawa couplings, hence some amount of flavour violation is
necessarily generated via quantum corrections in the charged lepton sector.

To calculate the minimum amount of lepton flavour violation in the charged
lepton sector, we assume that $Y^1_e$, $Y^2_e$ are diagonal at the cut-off scale 
$\Lambda>M_{\rm maj}$. Then, due to the radiative corrections from
the neutrino Yukawa couplings $Y^1_\nu$, $Y^2_\nu$, both charged lepton
Yukawa couplings become non-diagonal at the scale $M_{\rm maj}$. 
As discussed in the previous section, we now redefine the charged
lepton fields in order to bring the Yukawa coupling $Y^1_e$ into its
diagonal form. As a result, the off-diagonal elements 
of the charged lepton Yukawa coupling read, at the Majorana mass scale:
\begin{multline}
 \label{eq:LFV1}
 \left. (V_e^{L\dagger} Y^2_e)_{ij} \right|_{M_{\rm maj}} =
	\frac{\log\frac{\Lambda}{M_{\rm maj}}}{8 \pi^2}
	\big(- Y^1_\nu Y^{1\dagger}_\nu Y^2_e - Y^2_\nu Y^{1\dagger}_\nu Y^2_e (Y^1_e)^{-1} Y^2_e \\
		+Y^1_\nu Y^{2\dagger}_\nu Y^1_e  + Y^2_\nu Y^{2\dagger}_\nu Y^2_e \big)_{ij} \qquad \qquad \quad i<j\;.
\end{multline}
Below the Majorana mass scale the charged lepton Yukawa couplings
are also affected by the quantum effects, however the off-diagonal
elements at low energies are still given by the previous expression, 
up to second order effects. 

To estimate this contribution we assume $Y_e^2 = \xi_e Y_e^1$, with at the
cut-off scale. With this ansatz, eq.~(\ref{eq:LFV1}) reads:
\begin{equation}
 \left. (V_e^{L\dagger} Y^2_e)_{12} \right|_{M_{\rm maj}} =
	\frac{\log\frac{\Lambda}{M_{\rm maj}}}{8 \pi^2}
	(Y^1_\nu+\xi_e Y^2_\nu)_1(-\xi_e Y^{1*}_\nu+Y^{2*}_\nu)_2 Y^1_{e22}\;.
\end{equation}
Inserting this contribution into eq.~(\ref{eq:BRmuegamma}) one obtains
the approximate lower bound:
\begin{equation}
 \mathrm{BR} (\mu \rightarrow e\, \gamma) \gtrsim
  \frac{8 \alpha^3}{3 \pi^3}
  \left(\frac{\log\frac{\Lambda}{M_{\rm maj}}}{8 \pi^2}\right)^2
   |Y^1_{\nu1}+\xi_e Y^2_{\nu1}|^2 |Y^{2}_{\nu 2}-\xi_e^* Y^{1}_{\nu2}|^2 
   \left| f \!\! \left(\frac{m_t^2}{m_h^2} \right) \frac{|\lambda_6|v^2}{m_H^2} \right|^2 \;,
\end{equation}
which is saturated when the charged lepton Yukawa couplings are aligned
and when the Yukawa couplings of the heavy Higgs to the quarks are
negligible. 

To estimate the size of this lower bound, we will assume neutrino Yukawa 
couplings maximally misaligned with the form $Y^1_\nu = \frac{y_1}{\sqrt{2}}(0,1,1)^T$,
 $Y^2_\nu = \frac{y_2}{\sqrt{3}}(1,1,-1)^T$,
 being $y_1$ and $y_2$ the corresponding norms. With this choice, we obtain, in the limit 
$|\xi_e|\gg 1$ and $m_H\gg M_Z$ and taking $\Lambda = 10^4 M_{\rm maj}$,
\begin{equation}
 \mathrm{BR} (\mu \rightarrow e\, \gamma)\gtrsim 3\times 10^{-15} \times
    |\lambda_6 y_1 y_2 \xi_e^2|^2  \left(\frac{m_H}{3\;{\rm TeV}}\right)^{-4}\;.
\end{equation}
Given that this bound is very conservative, the observation of the
process $\mu\rightarrow e\gamma$ may be at the reach of the 
MEG experiment, which aims to 
$\mathrm{BR} (\mu \rightarrow e\, \gamma)>10^{-13}$~\cite{Maki:2008zz},
provided the couplings are sizable and provided the 
extra scalar degrees of freedom are not too heavy.

\section{Conclusions} 
\label{sec:conclusions}

We have considered in this paper an extension of the Standard Model
by one extra Higgs doublet and one or more Majorana right-handed neutrinos,
including in the Lagrangian all terms compatible with the Standard
Model gauge symmetry. We have calculated, using a renormalization group
approach, the quantum corrections to the neutrino parameters under the 
assumption that the right-handed Majorana mass scale is much larger than the 
mass of the Higgs mass eigenstates. We have argued that if the neutrino
Yukawa couplings are misaligned, the radiatively generated contribution
to the mass of the next-to-heaviest neutrino can be much larger than the tree
level mass. Furthermore, for reasonable choices of the parameters of the model,
the radiatively generated mass of the next-to-heaviest neutrino is a factor of a few
smaller than the mass of the heaviest neutrino. Since the mass hierarchy depends only
logarithmically on the masses of the extra degrees of freedom, this conclusion is
fairly insensitive to the scales at which the new physics appears.

Hence, in this simple model two puzzles in neutrino physics
can be simultaneously explained. First, the smallness of the neutrino masses
is explained by the see-saw mechanism. Secondly, the mild hierarchy between
the atmospheric and the solar neutrino mass scales is explained by the 
radiative origin of the mass of the next-to-heaviest neutrino, which is
suppressed by the loop factor but enhanced by the large logarithm of the
ratio between the heavy Majorana mass scale and the heavy Higgs scale. 
Furthermore, by making the heavy Higgs scale sufficiently large, 
all the successes of the Standard Model can be preserved, since all
low energy effects of the extended Higgs sector are suppressed at least
by two powers of the heavy Higgs mass.

The misalignment in the Yukawa couplings,  necessary for the radiative 
generation of the solar neutrino mass scale, amounts to new sources of
lepton flavour violation which also modify the structure of the leptonic
mixing matrix through the renormalization group evolution.
Therefore, we expect in this model deviations from the maximal 
atmospheric mixing and
from a vanishing $\theta_{13}$ due to quantum effects. 
We have carefully calculated these corrections and we have found that the 
radiatively generated angle  $\theta_{13}$ can be large enough to be measured
in present and future experiments.

\section*{Acknowledgements}

We are grateful to Carolin Br\"auninger, Alberto Casas,
Concha Gonz\'alez Garc\'ia, Thomas Hambye, 
Alejandra Melfo and Enrico Nardi for useful discussions. 
This work was partially supported by the DFG
cluster of excellence Origin and Structure of the Universe and by the Graduiertenkolleg
“Particle Physics at the Energy Frontier of New Phenomena”.
\appendix
\section{Appendix}
\subsection{Quantum corrections below $M_{\rm maj}$}

The one-loop $\beta$ functions of the multi-Higgs doublet model,
including the dimension five operator which yields neutrino masses, 
have been derived in \cite{Grimus:2004yh}.
The $\beta$ functions of the charged lepton Yukawa couplings, $Y_e^{a}$,
and the dimension five operators $\kappa^{ab}$ read, for energy
scales below the right-handed neutrino Majorana mass scale,
\begin{align}
 \beta_{Y_e^{a}}&=
	\left(-\frac{9}{4}g^2-\frac{15}{4}g'^2 \right) Y_e^{a}
	+\left[	
		3{\rm Tr}\left(Y_u^{a\dagger}Y_u^{c}+Y_d^{a}Y_d^{c\dagger}\right)
		+{\rm Tr}\left(Y_e^{a}Y_e^{c\dagger}\right)
	\right]Y_e^{c} \nonumber\\
	&\quad
	+Y_e^{a}Y_e^{c\dagger}Y_e^{c}
	+\frac{1}{2}Y_e^{c}Y_e^{c\dagger}Y_e^{a}\;,\\
 \beta_{\kappa^{ab}}&=
	\frac{1}{2} \left[ 
		Y_e^{c}Y_e^{c\dagger} \kappa^{ab}+
		\kappa^{ab} \left(Y_e^{c}Y_e^{c\dagger}\right)^T 
	\right]
	+2 \left[
		Y_e^{c}Y_e^{b\dagger} \kappa^{ac} +
		\kappa^{cb} \left( Y_e^{c}Y_e^{a\dagger} \right)^T 
	\right]\nonumber\\
	&\quad
	-2 \left[ 
		Y_e^{c}Y_e^{a\dagger} (\kappa^{cb} + \kappa^{bc}) +
		(\kappa^{ac}+\kappa^{ca}) \left( Y_e^{c}Y_e^{b\dagger} \right)^T 
	\right]\nonumber\\
	&\quad
	 + \left[ 3 {\rm Tr}(Y_u^{a}Y_u^{c\dagger} + Y_d^{a\dagger}Y_d^{c}) +
		{\rm Tr}(Y_e^{a\dagger}Y_e^{c})
			 \right] \kappa^{cb}\nonumber\\
	&\quad
	 +\kappa^{ac} \left[ 3 {\rm Tr}(Y_u^{b}Y_u^{c\dagger} + Y_d^{b\dagger}Y_d^{c})+
		{\rm Tr}(Y_e^{b\dagger}Y_e^{c})
			 \right]\nonumber\\
	&\quad
	 -3 g^2 \kappa^{ab}
	 +2 \lambda_{acbd} \kappa^{cd}\;,
\end{align}
where summation over repeated indices is understood and
the quartic couplings $\lambda$ are defined by
$V\supset \frac{1}{2}\lambda_{abcd}(\Phi^\dagger_a \Phi_b)(\Phi^\dagger_c \Phi_d)$.

In the case of only one right-handed neutrino and two Higgs doublets,
$\kappa^{11}(m_H)$ can be approximately written at the leading log at any mass scale $m_H < M_{\rm maj}$
in the form of eq.~\eqref{correction} which we repeat here for completeness:
\begin{align}
 \kappa^{11}(m_H) &\approx
	 \kappa^{11}(M_{\rm maj})
	-\frac{1}{16 \pi^2} \beta_{\kappa^{11}}(M_{\rm maj})\log\frac{M_{\rm maj}}{m_H}\\
	&\equiv \kappa^{11}(M_{\rm maj}) + B_{1a}\kappa^{a1}+\kappa^{1a}B_{1a}^T+ b \kappa^{22}\;,
\end{align}
where we have defined flavour matrices $B_{11}$, $B_{12}$ and the complex number $b$ by
\begin{align}
 \frac{16 \pi^2}{\log\frac{M_{\rm maj}}{m_H}} B_{11} &=
	  -\frac{1}{2} Y_e^{2}Y_e^{2\dagger} 
	+\frac{3}{2} Y_e^{1}Y_e^{1\dagger} 
	 - 3 {\rm Tr}(Y_u^{1}Y_u^{1\dagger} + Y_d^{1\dagger}Y_d^{1}) -
		{\rm Tr}(Y_e^{1\dagger}Y_e^{1}) 
	 + \frac{3}{2} g^2 
	 -\lambda_1\;,\\
\frac{16 \pi^2}{\log\frac{M_{\rm maj}}{m_H}}B_{12} &= 
	2 Y_e^{2}Y_e^{1\dagger}
	- 3 {\rm Tr}(Y_u^{1}Y_u^{2\dagger} + Y_d^{1\dagger}Y_d^{2}) -
		{\rm Tr}(Y_e^{1\dagger}Y_e^{2})
	- 2 \lambda_6\;,\\
\frac{16 \pi^2}{\log\frac{M_{\rm maj}}{m_H}} b &=	-2 \lambda_5\;.
\end{align}

\subsection{Quantum corrections above $M_{\rm maj}$}\label{sec:aboveMmaj}

In the case that the cut-off scale of the theory, $\Lambda$, is larger than
$M_{\rm maj}$, the relevant matricial couplings of the leptonic Lagrangian
are the charged lepton Yukawa couplings, $Y_e^{a}$, the neutrino 
Yukawa couplings, $Y_\nu^{a}$, and the right-handed Majorana mass matrix
$M_{\rm M}$. The corresponding $\beta$-functions are:
\begin{align}
  \beta^\Lambda_{Y_e^{a}}&= \beta_{Y_e^{a}} +{\rm Tr}(Y_\nu^{a\dagger}Y_\nu^{c}) Y_e^{c} 
	-2Y_\nu^{c}Y_\nu^{a\dagger}Y_e^{c} +\frac{1}{2}Y_\nu^{c}Y_\nu^{c\dagger}Y_e^{a} \label{beta:Ye}\;,\\
  \beta^\Lambda_{Y_\nu^{a}}&=
	\left[-\frac{9}{4}g^2-\frac{3}{4}g'^2 \right] Y_\nu^{a}
	+\left[	
		3{\rm Tr}(Y_u^{a}Y_u^{c\dagger}\!+\!Y_d^{a\dagger}Y_d^{c})
		+{\rm Tr}(Y_\nu^{a}Y_\nu^{c\dagger}+Y_e^{a\dagger}Y_e^{c})
	\right]Y_\nu^{c} \nonumber\\
	&\quad
	-2Y_e^{c}Y_e^{a\dagger}Y_\nu^{c}
	+Y_\nu^{a}Y_\nu^{c\dagger}Y_\nu^{c}
	+\frac{1}{2}Y_e^{c}Y_e^{c\dagger}Y_\nu^{a}
	+\frac{1}{2}Y_\nu^{c}Y_\nu^{c\dagger}Y_\nu^{a}\;,\\
  \gamma^\Lambda_{M_{\rm M}} &= - M_{\rm M}^{-1} \left[ \left( Y_\nu^{c\dagger}Y_\nu^{c}\right)^T M_{\rm M} + M_{\rm M} Y_\nu^{c\dagger}Y_\nu^{c} \right]\;.
 \end{align}

The running above $M_{\rm maj}$ modifies some of the expressions that we have
derived in this paper. The effects of the running can be easily incorporated
in our results by substituting
\begin{align}
 B_{1a} &\to B_{1a}+B^\Lambda_{1a}\;, & \kappa^{ab} \to \frac{Y_{\nu}^{a}  Y_{\nu}^{b\;T}}{M_{\rm maj}}\;.
\end{align}
More concretely,
the values for $B^\Lambda_{1a}$ read:
\begin{align}
 \frac{16 \pi^2}{\log\frac{\Lambda}{M_{\rm maj}}} B_{11}^\Lambda &= 
	\frac{9}{4} g^2 +\frac{3}{4}g'^2
	 - 3 {\rm Tr}(Y_u^{1}Y_u^{1\dagger} + Y_d^{1\dagger}Y_d^{1})
		- {\rm Tr}(Y_\nu^{1}Y_\nu^{1\dagger} +Y_e^{1\dagger}Y_e^{1})
	 \nonumber\\
	&\quad 
	-\frac{1}{2} Y_e^{2}Y_e^{2\dagger} 
	+\frac{3}{2} Y_e^{1}Y_e^{1\dagger} 
	-\frac{1}{2} Y_\nu^{1}Y_\nu^{1\dagger} 
	-\frac{1}{2} Y_\nu^{2}Y_\nu^{2\dagger}\;, \\
\frac{16 \pi^2}{\log\frac{\Lambda}{M_{\rm maj}}} B_{12}^\Lambda &= 
	2 Y_e^{2}Y_e^{1\dagger}
	- 3 {\rm Tr}(Y_u^{1}Y_u^{2\dagger} + Y_d^{1\dagger}Y_d^{2}) -
		{\rm Tr}(Y_e^{1\dagger}Y_e^{2}+Y_\nu^{1}Y_\nu^{2\dagger})\;.
\end{align}
Note that the running above $M_{\rm maj}$ does not modify the
value of $b$  nor $m_2$, cf.~eq.~\eqref{eq:m2}.

Besides, the running above $M_{\rm maj}$ also affects the structure
of $V_e^L$ at low energies. This can be taken into account with the
following substitution in the relevant formulas:
\begin{equation}
 \beta_{Y_e^1} \log \frac{M_{\rm maj}}{m_H} \to 
   \beta_{Y_e^1} \log \frac{M_{\rm maj}}{m_H} +
   \beta^{\Lambda}_{Y_e^1} \log \frac{\Lambda}{M_{\rm maj}}\;.
\end{equation}

\end{document}